\documentclass[aps,prl,twocolumn,showpacs,preprintnumbers,superscriptaddress,amsmath]{revtex4}
\usepackage{txfonts}
\usepackage{amssymb}
\usepackage{graphicx}
\usepackage{dcolumn}
\usepackage{bm}

\begin{document}

\title{Exciton-Plasmon-Photon Conversion in silver nanowire: polarization dependence}
\author{Lu-Lu Wang}
\author{Xi-Feng Ren}
\email{renxf@ustc.edu.cn.}
\author{Ai-Ping Liu}
\author{Liu Lv}
\author{Yong-Jing Cai}
\author{Guangcan Guo}
\author{Guoping Guo}
\address{Key Laboratory of Quantum Information, University of Science and Technology of China, Hefei
230026, People's Republic of China}
\begin{abstract}
Polarization dependence of the exciton-plasmon-photon conversion in
silver nanowire-quantum dots structure was investigated using a
scanning confocal microscope system. We found that the fluorescence
enhancement of the CdSe nanocrystals was correlated with the angle
between the excitation light polarization and the silver nanowire
direction. The polarization of the emission was also related with
the nanowire direction. It was in majority in the direction parallel
with nanowire.

\end{abstract}

\maketitle

Surface plasmon polaritons (SPPs), collective oscillating electrons
excited by electromagnetic field, have been studied for decades and
a great of interest has been injected into this area. SPPs are
involved in a wide range of phenomena \cite{Barnes03,Ozbay06},
including nanoscale optical waveguiding \cite{Bozhevolnyi06},
perfect lensing \cite{Pendry00}, extraordinary optical transmission
\cite{Ebbesen98}, subwavelength lithography \cite{Fang05}, and
ultrahigh sensitivity biosensing \cite{Liedberg83}. Analogous to
optical fiber in nanoscale, plasmonic waveguide has received more
and more attention since SPPs can propagates in nanoscale
structures. Among various plasmonic waveguides, silver nanowires
have some unique properties that make them particularly attractive,
such as low propagating loss due to their smooth surface and
scattering of plasmons to photons only at their sharp
ends\cite{sanders,ditlbacher}. Recently, not only the efficient
coupling method was proposed\cite{Knight,Pyayt,Dong}, but also the
correlation between incident and emission polarization in Ag
nanowire was studied\cite{xu1}.

Interaction of surface plasmons and quantum dots (QDs) attracted
more and more attention recently because of the importance of QDs in
quantum information \cite{hanson}. The metallic nanostructures can
be used to modify the emission properties of QDs. The enhancement or
quenching of QDs emission in various structures has been researched
in many works\cite{Fedutik,Matsuda,Ueda,Govorov,wu}. Recently, the
interaction of propagating SPs and QDs has also been investigated.
Akimov \emph{et al.} developed a single optical plasmons by coupling
the SPPs of silver nanowire with a single quantum dot\cite{lukin1}.
Wei \emph{et al.} showed the loss of energy and change of spectrum
when emission of QDs propagated along the silver
nanowires\cite{xu3}.

Polarization is an important property of photons, thus the
polarization properties of the exciton-plasmon-photon conversion of
QDs in the metallic nanowire should be studied intensively. In this
paper, we studied the polarization dependence of the efficiency of
exciton-plasmon-photon conversion in silver nanowire-quantum dots
structure. The results showed that the enhancement factor of QDs
fluorescence was correlated with the angle between the polarization
of the laser and the direction of the nanowire. Due to the antenna
effect of the nanowire, the polarization of the fluorescence was
also changed with the direction of the nanowire.

The 15 nm silica capped Ag nanowires (about 300 nm in diameter and
10 $\mu m$ in length) were prepared by using a wet chemistry method
described by Sun and Xia\cite{xia1,xia2}. This nanowire-QDs distance
was selected to get a high enhancement of fluorescence of
QDs\cite{Fedutik, lukin1}. A drop of solution of these Ag nanowires
in ethanol was deposited on a clean glass substrate and dried
naturally. Then a drop of CdSe/ZnS QDs (the fluorescence emission
wavelength centered at 610 nm) was deposited above the Ag nanowires
and covered with another glass slide immediately. The coverslip
should be lightly pressed in order to get uniform distribution of
the QDs. After the sample was dry, the capped glass slide was
abandoned while the silver nanowires covered with QDs on the
substrate were used for the following measurement.

\begin{figure}[b]
\includegraphics[width=8.0cm]{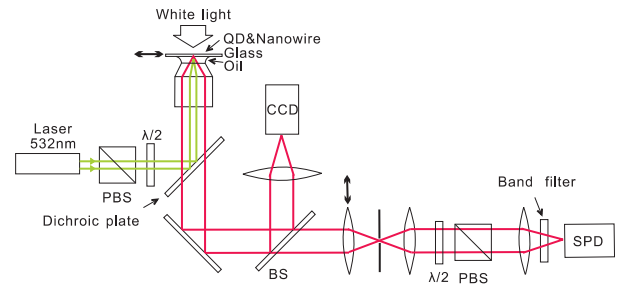}
\caption{(Color online) Sketch of the experimental setup. A homemade
scanning confocal microscope was used to investigate the
polarization dependence of exciton-plasmon-photon conversion in
silver nanowire-QDs structure.}
\end{figure}

The experimental setup was shown in Fig. 1. The sample placed on a
three-dimensional piezoelectric transition (PZT) stage were excited
by a continuous-wave laser at 532 nm, whose polarization can be
adjusted by a $\lambda/2$ plate (work wavelength 532 nm) behind a
polarization beam splitter(PBS). The laser light was focused on the
sample by a 60$\times$ oil objective (NA = 1.35), and the
fluorescence from the QDs were collected with the same objective and
detected by a single photon detector(SPD, SPCM-15, PerkinElmer
Optoelectronics, Canada). A green-orange dichroic plate was used to
reflect the laser light and transmit the fluorescence from QDs(the
emitting peak is centered at 610 nm at room temperature). Spatial
filtering was realized by the confocal lens with a pinhole(diameter
30 $\mu m$) in center. This setup can collect the light from an area
about 1.5 $\mu m$ in diameter on sample surface. The area can be
selected artificially by moving the the front lens of the confocal
lens system. A narrow band filter(center wavelength 633nm) was also
used to reflect the illuminated light before the SPD. The
polarization of the fluorescence was analyzed by another $\lambda/2$
plate (work wavelength = 633 nm) combined with a PBS. A CCD camera
was used for direct imaging.

\begin{figure}[b]
\includegraphics[width=6.0cm]{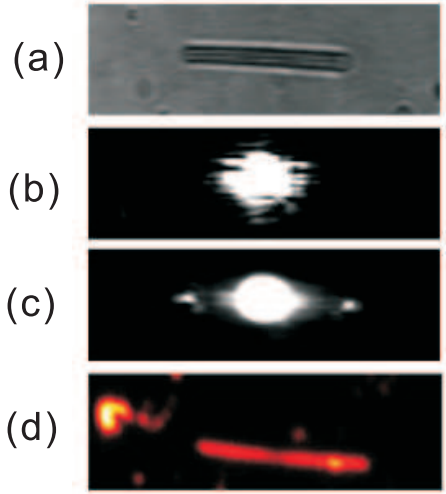}
\caption{(Color online) (a) CCD image of a silver nanowire.(b) Image
with 532 nm laser focused on the middle of the nanowire. (c) Image
of the QDs emission as the same setup as case b. (d) Scanning image
of the QDs emission.}
\end{figure}

Fig. 2(a) gave the CCD picture of a Ag nanowire with length about 14
$\mu m$. When the 532 nm laser was focused on the middle of the
nanowire, the ends of nanowire kept dark(Fig. 2(b)), which means
that surface plasmons cannot be excited in the nanowire. The reason
was that the momentum difference between the surface plasmons and
photons was not bridged due to the smooth surface of nanowire. This
phenomenon was observed in many works\cite{sanders,ditlbacher}.
However, the observed phenomenon changed as the nanowire was coated
by the QDs. When using a $610\pm17 nm$ band-pass filter in front of
the CCD to block the laser light, we can see the fluorescence on the
ends of nanowire obviously(Fig. 2(c)). Contrast with Fig. 2(b), we
considered that the fluorescence light at the ends of nanowire were
came from the propagating SPPs excited by QDs' fluorescence around
the nanowire but not the 532 nm laser. We also scanned the sample in
the area of $30 \times 10  \mu$m (the scan step is 0.2 $\mu$m)
around the Ag nanowire and collected the fluorescence. The result
was shown in Fig. 2(d). Strong fluorescence near the nanowire can be
clearly observed. The bright spot outside the nanowire in Fig. 2(d)
was attributed to the enhancement of nanoparticle doping.

\begin{figure}[b]
\includegraphics[width=8.0cm]{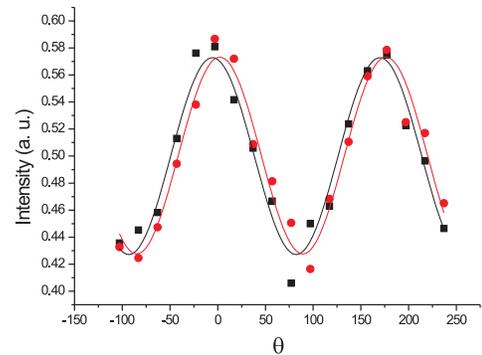}
\caption{(Color online) Intensity of fluorescence from the
middle(black square dots) and the right end of the nanowire(red
round dots) as the function of $\theta$. The results were nicely
fitted with cosine function.}
\end{figure}

The relation between the intensity of the fluorescence and the cross
angle of silver nanowire and laser light polarization was
investigated by rotating the first $\lambda/2$ plate (532 nm). By
moving the front lens of the confocal system, we can selectively
collect the fluorescence from the middle part(excitation spot) or
one end of the nanowire. For both cases, different counts were
recorded by the SPD for various polarization of the excitation
light, as shown in Fig. 3. $\theta$ was defined as the cross angle
between the silver nanowire and the polarization of excitation
laser. The normalized counts from the middle and right end of the Ag
nanowire were nearly coincident, both of them came to maximum when
the polarization of laser was parallel with the nanowire, and to
minimum when they were perpendicular. They were nicely fitted with
cosine function of $\theta$. $I_{max}/I_{low}$ were about 1.34 for
the both cases of the middle part and the right end, where $I_{max}$
and $I_{low}$ were the maximum and minimum counts respectively. The
same values supported that the fluorescence from the end of the
nanowire came from propagating SPs excited by the QDs near the Ag
nanowire. This polarization dependence may be due to the different
excitation efficiency of the SPPs, which contributed to the
enhancement of the fluorescence\cite{Fedutik,Zou}.
\begin{figure}[b]
\includegraphics[width=8.0cm]{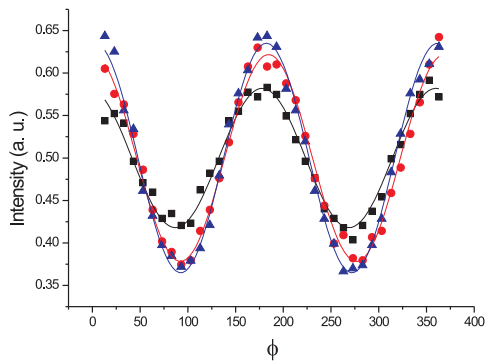}
\caption{(Color online) Intensity of fluorescence from the middle of
the nanowire as the function of $\phi$. For black square dots, red
round dots and blue triangle dots, $\theta$ were 0, 45 and 90 degree
respectively.}
\end{figure}

\begin{figure}[b]
\includegraphics[width=8.0cm]{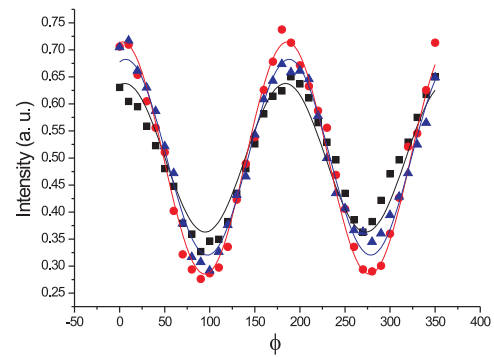}
\caption{(Color online) Intensity of fluorescence from the right end
of the nanowire as the function of $\phi$. For black square dots,
red round dots and blue triangle dots, $\theta$ were 0, 45 and 90
degree respectively. }
\end{figure}
The polarization of the fluorescence from the middle of the
nanowire(excitation spot) and one end of the nanowire were also
analyzed. By rotating another $\lambda/2$ plate (633 nm) before the
SPD, we can selectively detected the fluorescence in different
polarization. Three cases for the excitation laser polarization with
$\theta = 0^\circ$, $45^\circ$ and $90^\circ$ were discussed as
shown in FIG. 4. $\phi$ was defined as the angle between Ag nanowire
and polarization of detected fluorescence. In all the cases, the
fluorescence showed maximum or minimum counts as the detected
polarization was parallel or perpendicular with the nanowire
respectively. It means that the fluorescence was mainly polarized
towards the direction of nanowire. This phenomenon may came from the
antenna effect of the metal nanowire structure, which was similar
with the result of $\lambda/2$ dipole antennas in ref. \cite{curto}.
The polarization ratio $I^{mid}_{\|}/I^{mid}_{\bot}$ is 1.39
($\theta = 0^\circ$), 1.65 ($\theta = 45^\circ$), 1.74 ($\theta =
90^\circ$) respectively, where the $I_{\|}$($I_{\bot}$) was the
intensity of fluorescence with the polarization
parallel(perpendicular) with the nanowire. The similar result was
found for the fluorescence from the right end of the nanowire, as
shown in Fig. 5, while the polarization ratio
$I^{end}_{\|}/I^{end}_{\bot}$ was changed to 1.76 ($\theta =
0^\circ$), 2.01 ($\theta = 45^\circ$), 2.13 ($\theta = 90^\circ$).
They were all higher than the previous case, which may be attributed
to the polarization selection in the propagation of SPPs\cite{xu1}.

Several nanowires were investigated using the same method and shown
the similar results. The only differences were the unequal
polarization ratio $I_{max}/I_{low}$ and $I_{\|}/I_{\bot}$, which
were related to the length and diameter of nanowires. As a
comparison, we also studied the cases of pure QDs sample using the
same setup. It was found that not only the intensity of the
fluorescence was independent of the polarization of the excitation
light, but also the polarization of the fluorescence was random.
This was quite different with the silver nanowire-quantum dots
system, consequently, the observed results shown in the Figs. 3-5
were caused by the presence of the SPPs of the nanowire.

In summary, we investigated the polarization dependence in
exciton-plasmon-photon conversion in silver nanowire-quantum dots
system. Due to the effect of the SPPs, the whole process was quite
different from the excitation of the pure QDs. Besides the strong
correlation between the fluorescence enhancement factor and the
polarization of the excitation laser, the polarization of the
fluorescence itself was also related with the direction of the
nanowire. Our results will be useful in the further research on the
interactions of SPs and QDs.

The authors thank Chang-Ling Zou and Jin-Ming Cui for useful
discussion. This work was funded by the National Basic Research
Programme of China (Grants No.2011CBA00200), the Innovation funds
from Chinese Academy of Sciences, and the National Natural Science
Foundation of China (Grants No.10904137), Anhui Provincial Natural
Science Foundation(Grants No. 090412053) and Science and
Technological Fund of Anhui Province for Outstanding Youth(Grants
No.2009SQRZ001ZD).

\end{document}